\documentstyle[pra,aps,twocolumn,epsfig]{revtex}

\begin{document}

\draft

\title{Homodyne Bell's inequalities for entangled
mesoscopic superposition states}

\author{Andrzej Dragan}

\address{Instytut Fizyki Teoretycznej, Uniwersytet Warszawski, Ho\.{z}a 69,
PL--00--681 Warszawa, Poland}

\author{Konrad Banaszek}

\address{Center~for~Quantum~Information and
Rochester~Theory~Center for Optical Science and Engineering,\\
University~of~Rochester, Rochester NY 14627}

\date{\today}

\maketitle

\begin{abstract}
We present a scheme for demonstrating the violation of Bell's
inequalities using a a spin-1/2 system entangled with a pair of
classically distinguishable wave packets in a harmonic potential. For
electromagnetic fields, such wave packets can be represented by coherent
states of a single radiation mode. The proposed scheme involves standard
spin-1/2 projections and continuous measurements of the position and the
momentum of the harmonic oscillator system, which for a radiation mode
can be realized by means of homodyne detection. We discuss effects of
imperfections, including non-unit efficiency of the homodyne detector,
and point out a direct link between the visibility of interference and
the violation of Bell's inequalities in the described scheme.
\end{abstract}


\section{Introduction}

Quantum mechanical superposition principle brings unexpected
and counterintuitive consequences when applied to macroscopic
systems. Presumably the most famous example is Schr\"{o}diger's cat,
which remains half-alive and half-dead while entangled with a decaying
radioactive atom \cite{Erwin}. Recent experimental advances have opened
up new possibilities to study the superposition principle beyond purely
microscopic domain. It is now possible to produce in a laboratory states
of the type:
\begin{equation}
\label{Eq:Kot}
|{\cal K}\rangle=\frac{1}{\sqrt{2}}(|\!\uparrow\,\rangle\otimes
|\alpha\rangle
+|\!\downarrow\,\rangle \otimes |\!-\!\alpha\rangle)
\end{equation}
where $|\!\uparrow\,\rangle$, $|\!\downarrow\,\rangle$ are two orthogonal
states of a spin-1/2 system, and $|\alpha\rangle$, $|\!-\!\alpha\rangle$
are two distinguishable coherent wave packets of a harmonic
oscillator. Such states have been generated for a trapped ion \cite{Ion}
and a microwave cavity field entangled with an atom \cite{CavityQED}. They
can be considered as mesoscopic equivalents of the example used by
Schr\"{o}dinger in his original argument. The entangled states defined
in Eq.~(\ref{Eq:Kot}) are closely related to the issue of generating
and detecting coherence between classically distinguishable states
\cite{Koty,Savage,Genovese}.

In this paper we show how states described by Eq.~(\ref{Eq:Kot})
can be used to test incompatibility of quantum mechanics with local
realism. Specifically, we derive Bell's inequalities
which are violated by the Schr\"{o}dinger cat states. These inequalities
are based on the continuous measurements of position and momentum
observables for the harmonic oscillator subsystem, and standard
projections for the spin-1/2 subsystem. An interesting feature of our
proposal is that detection of continuous variables having a well-defined
classical analog allows one to investigate the macroscopic limit
of violating Bell's inequalities by the entangled states given
by Eq.~(\ref{Eq:Kot}). We demonstrate that in the limit of large
wave packet amplitudes
a substantial violation of Bell's inequalities is possible, provided
ideal noise-free detection and lack of decoherence. We also perform
a general analysis of the proposed scheme, including imperfect detection and
dissipation, which gives a quantitative description of the disappearance
of nonlocal phenomena in the presence of these deleterious effects.
In particular, our analysis shows that the violation of Bell's
inequalities in the proposed scheme vanishes at the same rate the
visibility of interference between the two distinct wave packets given
by the states $|\alpha\rangle$ and $|\!-\!\alpha\rangle$. This result
illustrates the close link between nonlocality and quantum coherence.
As it will be clear from the following calculations, all these features
are universal, i.e.\ they are independent of the particular form
of the wave packets involved in the superposition. We also point out
that in order to demonstrate the violation of Bell's inequality,
only one of the two measurements applied to the
harmonic oscillator subsystem needs to have microscopic resolution, whereas
the second one is relatively insensitive to losses. Thus, our scheme provides
another example of a situation described by Yurke and Stoler
\cite{YurkStolPRL97}, whose proposal for observing the violation
of local realism employed a combination of
sensitive and insensitive detectors.

For concreteness, we shall consider here a quantum optical realization
of the entangled states $|{\cal K}\rangle$. In the case of a radiation mode,
which will serve in this paper as a physical realization of the
harmonic oscillator subsystem, position and momentum correspond
to a pair of quadratures, which can be measured with the help of
homodyne detection \cite{HomodyneBell}. The imperfect measurement
of quadratures can be described by an efficiency parameter
$\eta$. This parameter can be straightforwardly generalized to
include interaction of the electromagnetic field with an external
environment, which serves as a standard model for decoherence
\cite{WheelerZurek,WallMilbPRA85,CatReview}. The quantum optical context
will be used here just to fix the notation, and our
calculations retain validity for an arbitrary physical realization
of a spin-1/2 particle entangled with a harmonic oscillator system.

This paper is organized as follows. First, in Sec.~\ref{Sec:Heuristic},
we present a simple heuristic idea behind the construction
of Bell's inequalities for mesoscopic superposition
states. This idea is elucidated in quantitative terms in
Sec.~\ref{Sec:Quantitative}. Sec.~\ref{Sec:Violation} discusses the
violation of Bell's inequality including the realistic case of losses,
and Sec.~\ref{Sec:Experimental} briefly reviews some of the experimental
aspects.  Finally, Sec.~\ref{Sec:Conclusions} concludes the paper.

\section{Heuristic considerations}
\label{Sec:Heuristic}

\begin{figure}
\includegraphics{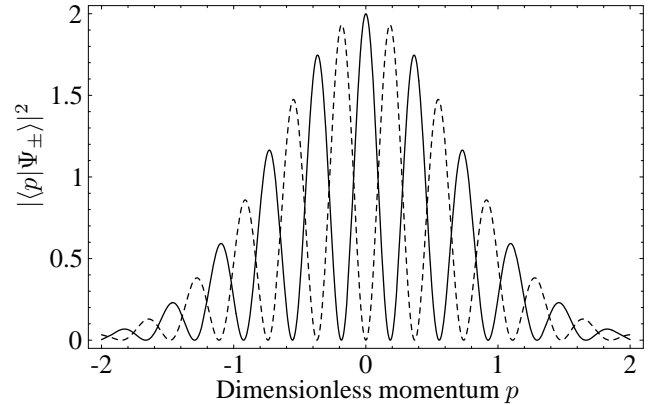}
\caption[t]{\label{wykres1}
Plot of the momentum distributions for the superpositions $|\Psi_+\rangle$
(solid line) and $|\Psi_-\rangle$ (dashed line), assuming perfect
noise-free measurement and $\alpha=6$. Location of maxima for one state
corresponds to the location of minima for the other one.}
\end{figure}

We shall start from giving a simple heuristic argument which motivated
us to formulate Bell's inequalities for Schr\"{o}dinger cat states. For
concreteness, we shall assume that the states $|\alpha\rangle$
and $|\!-\!\alpha\rangle$ describe two Gaussian wave packets in a
harmonic potential centered around dimensionless positions $\sqrt{2}\alpha$
and $-\sqrt{2}\alpha$, respectively, with zero average momentum and ground
state widths.  From the following discussion it will be clear
that the violation of Bell's inequalities in our scheme is completely
insensitive to the specific form of the wave packets.  The states
$|\alpha\rangle$ and $|\!-\!\alpha\rangle$ have the scalar product equal
to $\langle-\alpha|\alpha\rangle=\exp(-2\alpha^2)$, and for sufficiently
large $\alpha$ they can be considered as approximately orthogonal. In
this case, one can establish a formal analogy \cite{KW}
between the state $|{\cal
K}\rangle$ and the singlet state of two spin-1/2 particles used in
original Bell's argument, based on the correspondence:
$|\alpha\rangle \mapsto |\!\downarrow\,\rangle$ and $|\!-\!\alpha\rangle
\mapsto - |\!\uparrow\,\rangle$. Following this analogy, in order
to violate Bell's inequalities one should be able to perform two
noncommuting measurements in the subspace spanned by $|\alpha\rangle$
and $|\!-\!\alpha\rangle$, which would correspond to projecting the
spin onto two different directions.  As the first measurement on the
harmonic oscillator subsystem, let us simply choose the projection in
the basis $\{ |\alpha\rangle, |\!-\!\alpha\rangle\}$. This measurement
can be effectively accomplished by the measurement of position: if the
distance between the centers of the wave packets is much larger than their
spatial extent, the sign of the position variable almost unambiguously
discriminates between the states $|\alpha\rangle$ and $|-\alpha\rangle$.
As the second measurement on the harmonic oscillator subsystem we shall
take the projection in the basis of the superpositions:
\begin{equation}
\label{Eq:Psipm}
|\Psi_\pm\rangle=\frac{1}{\sqrt{N_{\pm}}}
(|\alpha\rangle\pm |\!-\!\alpha\rangle),
\end{equation}
where $N_{\pm} = 2(1 \pm e^{-2\alpha^2})$ are the normalization constants.
As depicted in Fig.~\ref{wykres1}, these two superpositions generate
distinct interference patterns in the momentum distribution: location
of the maxima for the state $|\Psi_+\rangle$ corresponds to the minima for
the state $|\Psi_-\rangle$, and vice versa. Consequently, we can
approximately discriminate between the superpositions $|\Psi_+\rangle$
and $|\Psi_-\rangle$ by checking whether the result of the momentum
measurement falls within the vicinity of the interference fringes
either for the state $|\Psi_+\rangle$ or $|\Psi_-\rangle$. Of course,
such discrimination is imperfect as the momentum distributions partially
overlap; nevertheless, we shall demonstrate that the error rate involved
is low enough to enable the violation of Bell's inequalities.

\section{Quantitative analysis}
\label{Sec:Quantitative}

Let us now discuss the idea sketched above in quantitative terms.
As a concrete physical realization, we will take $|\alpha\rangle$ and
$|\!-\!\alpha\rangle$ to be two coherent states of a single radiation mode,
with $\alpha$ assumed to be real. The optical analog of the position
and momentum observables is a pair of canonically conjugated quadratures
defined in general as $\hat{x}_\theta = (e^{i\theta} \hat{a}^\dagger +
e^{-i\theta} \hat{a})/\sqrt{2}$. Two values of the phase: $\theta =
0$ and $\theta = \pi/2$ correspond to the position and the momentum
operators, respectively.  The standard technique for measuring quadratures
is homodyne detection, described by the positive operator-valued measure
\cite{KiK}:
\begin{equation}
\label{Kwadraturka}
\hat{\cal H}(x;\theta) = \frac{1}{\sqrt{\pi(1-\eta)}}
\exp\left(-\frac{(x/\sqrt{\eta}-\hat{x}_\theta)^2}{1/\eta-1}\right),
\end{equation}
where $\eta$ is the detection efficiency.
More generally, the parameter $\eta$ can include dissipation of
the electromagnetic field generated by an interaction with an environment.
The perfect noise-free
measurement of quadratures is obtained in the limit $\eta=1$, whereas
$\eta < 1$ describes non-ideal detection. Such non-ideal
detection corresponds to a blurred
measurement of position and momentum with finite resolution equal
to $\sqrt{(1/\eta -1)/2}$, expressed
in the canonical dimensionless units of the
harmonic oscillator.

The states $|\Psi_{\pm}\rangle$ defined in
Eq.~(\ref{Eq:Psipm}) generate the following interference patterns for
$\theta=\pi/2$:
\begin{eqnarray}
\label{Rozklady}
\lefteqn{
\langle \Psi_{\pm} |
\hat{\cal H}(x; {\textstyle \frac{\pi}{2}})
| \Psi_{\pm} \rangle} 
\nonumber \\
& = &
\frac{2}{\sqrt{\pi} N_{\pm}}
e^{-x^2}[1\pm e^{-2 (1-\eta)\alpha^2}\cos (\sqrt{8\eta}\alpha x)].
\end{eqnarray}
It is easily seen that within the Gaussian envelope given by the
factor $e^{-x^2}$, the spacing between the interference fringes is
given by $T=\pi/\sqrt{2\eta}\alpha$. For the state $|\Psi_+\rangle$ the
interference pattern has maxima for integer multiples of $T$, whereas the
interference maxima for the state $|\Psi_-\rangle$ are shifted by $T/2$.

With the above definitions in hand, we can now specify the
measurement scheme which leads to the violation of Bell's inequalities
for Schr\"{o}dinger cat states. The party measuring the spin-1/2
subsystem performs standard spin projection along a direction selected
randomly between two unit vectors ${\bf a}= (a_x,a_y,a_z)$ or ${\bf
a}' =(a_x',a_y',a_z')$. The party measuring the harmonic oscillator
subsystem applies homodyne detection with the phase $\theta$ adjusted
to either $0$ or $\pi/2$. For $\theta = 0$, this realizes the quantum
optical analog of the position measurement. In this case, the continuous
outcome $x$ of homodyne detection needs to be converted into its sign,
as the regions $x>0$ and $x<0$ correspond respectively to detecting the
states $|\alpha\rangle$ and $|-\alpha\rangle$. Consequently for $\theta =
0$ we effectively perform the measurement of the following operator:
\begin{equation}
\hat{\cal C}_0 = \int_{x>0} dx \, \hat{\cal H}(x;0)
- \int_{x<0} dx \,  \hat{\cal H}(x;0).
\end{equation}
For the setting $\theta = \pi/2$, it is necessary to discriminate between
the location of fringes for the superpositions $|\Psi_+\rangle$ and
$|\Psi_-\rangle$. For this purpose, let us define two disjoint subsets
$\Lambda_+$ and $\Lambda_-$
of the possible outcomes of the homodyne measurement:
\begin{equation}
\Lambda_{\pm} = \bigcup_{n=-\infty}^{\infty} [(n \mp 1/4)T,(n+1/2 \mp
1/4 )T].
\end{equation}
These two subsets are obtained by comparing which of the two values
computed in Eq.~(\ref{Rozklady}):
$\langle \Psi_{+} |
\hat{\cal H}(x; {\textstyle \frac{\pi}{2}})
| \Psi_{+} \rangle$ or
$\langle \Psi_{-} |
\hat{\cal H}(x; {\textstyle \frac{\pi}{2}})
| \Psi_{-} \rangle$
is
larger for each $x$, assuming that the normalization factors $N_{\pm}$
are approximately equal.

The results $+1$ and $-1$ are assigned to the measured quadratures $x$
which belong respectively to the subsets $\Lambda_+$ and $\Lambda_-$. Thus
we measure the operator:
\begin{equation}
\hat{\cal C}_{\pi/2} = \int_{\Lambda_{+}} dx \,  
\hat{{\cal H}}(x; {\textstyle \frac{\pi}{2}})
- \int_{\Lambda_{-}} dx \,  
\hat{{\cal H}}(x; {\textstyle \frac{\pi}{2}}).
\end{equation}
The measurements performed on the harmonic oscillator subsystem are
correlated with the standard spin measurement along directions defined
by the unit vectors ${\bf a}$ and ${\bf a}'$. Thus we are interested in
the correlation functions of the form:
\begin{equation}
\label{Korel}
E({\bf a}, \theta) = \langle {\cal K} | 
({\bf a} \cdot
\hat{\mbox{\boldmath $\sigma$}})
\otimes \hat{\cal C}_\theta | {\cal K} \rangle
\end{equation}
where $\theta$ stands for $0$ or $\pi/2$. The outcomes of the measurements
performed both on the spin-1/2 and on the harmonic oscillator subsystems
correspond to local realities bounded by $-1$ and $1$. Consequently,
any of Bell's inequalities derived for a pair of spin-1/2 projections
can be used to test local reality in our measurement
scheme. We shall consider the following Bell combination constructed
from four correlation functions \cite{Bell}:
\begin{equation}
\label{BellComb}
S = E({\bf a}, 0) + E({\bf a}, {\pi/2})
+ E({\bf a}', 0) - E({\bf a}', {\pi/2}).
\end{equation}
For local hidden variable theories, the absolute value of this combination
is bounded by $|S| \le 2$. Explicit calculation of the combination $S$
in our scheme is simplified by the following symmetries of the operators
$\hat{{\cal C}}_0$ and $\hat{{\cal C}}_{\pi/2}$:
\begin{eqnarray}
& &
\langle \alpha | \hat{\cal C}_0 |\alpha \rangle =
- \langle -\alpha | \hat{\cal C}_0 | \!-\!\alpha \rangle
\nonumber \\
& &
\langle \alpha | \hat{\cal C}_0 | \!-\! \alpha \rangle
= 
\langle - \alpha | \hat{\cal C}_0 | \alpha \rangle = 0
\nonumber \\
& &
\langle \alpha | \hat{\cal C}_{\pi/2} | \alpha \rangle
= 
\langle -\alpha | \hat{\cal C}_{\pi/2} | \!-\!\alpha \rangle.
\end{eqnarray}
With the help of these identities, one easily obtains the Bell combination
expressed in terms of the matrix elements of the operators $\hat{\cal
C}_0$ and $\hat{\cal C}_{\pi/2}$:
\begin{eqnarray}
S & = & 
(a_x-a_x') \text{Re}
\langle \alpha | \hat{\cal C}_{\pi/2} |\!-\!\alpha \rangle
\nonumber \\
& &
+
(a_y-a_y') \text{Im}
\langle \alpha | \hat{\cal C}_{\pi/2} |\!-\!\alpha \rangle
+
(a_z+a_z')\langle \alpha | \hat{\cal C}_0 |\alpha \rangle.
\nonumber \\
\label{BellCombBIS}
& &
\end{eqnarray}
It is seen that there are two matrix elements relevant to the Bell
combination: the diagonal element of the operator $\hat{\cal C}_0$
which can be easily expressed in terms of the error function:
\begin{equation}
\label{elementy}
\langle \alpha | \hat{\cal C}_0 |\alpha \rangle  = 
\text{erf}\,( \sqrt{2\eta}\alpha)
\end{equation}
and the off-diagonal element of the operator $\hat{\cal C}_{\pi/2}$:
\begin{eqnarray}
\langle \alpha | \hat{\cal C}_{\pi/2} |\! -\!\alpha \rangle
& = & -e^{-2\alpha^2}+\frac{2}{\sqrt{\pi}}e^{-2\alpha^2(1-\eta)}\times
\nonumber \\
& &
\times\sum_{n=-\infty}^{\infty}\int_{(n-1/4)T}^{(n+1/4)T}e^{-x^2}\cos (\sqrt{8\eta}\alpha x)dx.
\nonumber \\
\label{OffCpi2}
& &
\end{eqnarray}
In the following, we will perform a detailed analysis of these two matrix
elements, and discuss the violation of Bell's inequalities which can be
achieved in the presented scheme.

\section{Violation of Bell's inequality}
\label{Sec:Violation}

In order to discuss the violation of Bell's inequality for the combination
$S$, let us first perform maximization over the unit vectors ${\bf a}$
and ${\bf a}'$ along which the spin-1/2 system is measured. An easy
calculation shows that the maximum value of $S$ reads:
\begin{equation}
\label{EsMax}
S_{\text{max}}=2\sqrt{\langle \alpha |
\hat{\cal C}_0 |\alpha \rangle^2+|\langle \alpha
| \hat{\cal C}_{\pi/2} |\! -\!\alpha \rangle|^2}
\end{equation}
and that it is obtained for the following directions of the spin
measurements:
\begin{eqnarray}
& &
a_x=-a_x' = \frac{2}{S_{\text{max}}}
\text{Re} \,
 \langle \alpha | \hat{\cal C}_{\pi/2} | \!-\!\alpha \rangle
\nonumber \\
& &
a_y= -a_y' = \frac{2}{S_{\text{max}}}
\text{Im} \,
 \langle \alpha | \hat{\cal C}_{\pi/2} | \!-\!\alpha \rangle
\nonumber \\
& &
a_z=~~a_z'~ = \frac{2}{S_{\text{max}}}
 \langle \alpha | \hat{\cal C}_0 | \alpha \rangle.
\end{eqnarray}

\begin{figure}
\includegraphics{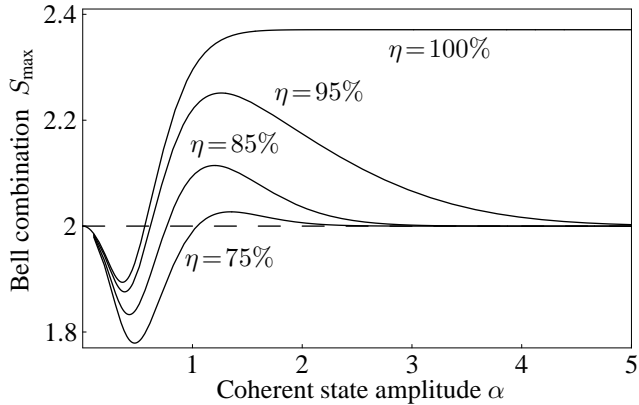}
\caption[]{\label{wykres2}
Maximum violation of Bell's inequality for Schr\"{o}dinger cat states
as a function of the coherent state amplitude $\alpha$. The graphs
correspond to measurements performed using a homodyne detector having
several different efficiencies.}
\end{figure}

In Fig.~\ref{wykres2} we plot $S_{\text{max}}$ as a function of $\alpha$
for several values of the detection efficiency $\eta$. In the case of
perfect detection of quadratures, the value of $S_{\text{max}}$ tends with
increasing $\alpha$ to a constant value, equal about 2.37. This result
clearly contradicts predictions of local hidden variable theories. In
the case of non-ideal measurement of quadratures, the violation of Bell's
inequality can be still observed for sufficiently high  efficiency $\eta$
of homodyne detection, but this effect vanishes with the increasing
coherent state amplitude $\alpha$.

In order to understand the behavior of $S_{\text{max}}$ in simple
terms, we will now perform an approximate analysis of the expression
(\ref{EsMax}) valid for large $\alpha$. Under an additional condition
$\sqrt{\eta}\alpha \gg 1$, which means that homodyne detection with the
phase $\theta=0$ is capable of discriminating between the wave packets
$|\alpha\rangle$ and $|-\alpha\rangle$, we have:
\begin{equation}
\label{Eq:C0Approx}
\langle \alpha | \hat{\cal C}_0 |\alpha \rangle \approx 1.
\end{equation}
The approximately constant value of this matrix element means that
the outcome of the homodyne measurement performed
for $\theta=0$ is insensitive to the detection efficiency. 

Approximation of the matrix element
$\langle\alpha| \hat{C}_{\pi/2} | - \alpha\rangle$ 
given in Eq.~(\ref{OffCpi2}) is
slightly more intricate. Large $\alpha$ means that the spacing $T$
between the interference fringes observed in the momentum distribution
is small compared to the extent of the wave packets. Consequently,
we can assume that the Gaussian envelope multiplying the integrand in
Eq.~(\ref{OffCpi2}) is constant over each of the integration intervals,
which allows us to evaluate the integrals analytically. Furthermore,
in this regime the sum over $n$ of  the remaining Gaussian factors can
be approximated by an integral. Thus we obtain:
\begin{eqnarray}
\lefteqn{
\sum_{n=-\infty}^{\infty}
\int_{(n-1/4)T}^{(n+1/4)T} e^{-x^2}\cos (\sqrt{8\eta}\alpha x)dx
}
 & &
\nonumber \\
& \approx  & \frac{1}{\pi} \sum_{n=-\infty}^{\infty} T e^{-(n T)^2}
\approx \frac{1}{\sqrt{\pi}}.
\end{eqnarray}
This expression yields the following approximate formula for the
off-diagonal matrix element of the operator $\hat{\cal C}_{\pi/2}$:
\begin{equation}
\label{Eq:Cpi2Approx}
\langle \alpha | \hat{\cal C}_{\pi/2} |\! -\!\alpha \rangle  
\approx \frac{2}{\pi} \exp[-2\alpha^2 (1-\eta)],
\end{equation}
where we have made use of the assumption $\sqrt{\eta} \alpha \gg 1$
in order to eliminate the first term from Eq.~(\ref{OffCpi2}).
This matrix element depends critically on
the efficiency of the homodyne detector.

Thus we
finally arrive to the following expression for the Bell combination:
\begin{equation}
\label{Eq:Smaxapprox}
S_{\text{max}} \approx 2\sqrt{1+
\left(\frac{2}{\pi}\exp[-2\alpha^2 (1-\eta)]\right)^2}.
\end{equation}
In the case of perfect homodyne detection, the right hand side of the
above formula is constant and equal to $2\sqrt{1+4/\pi^2}$. This is
the asymptotic value observed in Fig.~\ref{wykres2} in the plot of
$S_{\text{max}}$ for $\eta=100\%$. For imperfect homodyne detection,
the violation of Bell's inequality is damped for large $\alpha$ by the
exponential factor $\exp[-2\alpha^2 (1-\eta)]$, which decreases with
the increasing separation between the positions of the wave packets.
Let us note that the calculations which led to the approximate form
of the matrix elements given in Eqs.~(\ref{Eq:C0Approx}) and
(\ref{Eq:Cpi2Approx}) are independent of the particular form of
the wave packets $|\alpha\rangle$ and $|-\alpha\rangle$ involved
in the superposition $|{\cal K}\rangle$. In order to derive
Eq.~(\ref{Eq:C0Approx}) one only needs to assume that the spatial
extent of the wave packet is smaller that the separation between
them. Similarly, Eq.~(\ref{Eq:Cpi2Approx}) follows directly
from the assumption that the wave packet envelopes in the momentum
domain vary slowly on the scale of the oscillations generated
by the interference term.

It is interesting to note that the violation of Bell's inequality in the
proposed scheme is directly related to the visibility of interference
between the wave packets $|\alpha\rangle$ and $|\!-\!\alpha\rangle$. Of
course, homodyne detection performed alone on the harmonic oscillator
subsystem does not reveal any interference, as its reduced density
matrix is just a statistical mixture of $|\alpha\rangle$ and
$|\!-\!\alpha\rangle$. However, conditioning the homodyne measurement
on a specific outcome of the spin projection yields a clear signature
of interference. A simple calculation gives the following probability
distribution of obtaining the quadrature $x$ for the phase $\theta =
\pi/2$ conditioned on the spin up outcome for a projection along the
axis ${\bf a}$:
\begin{eqnarray}
p(x; {\textstyle \frac{\pi}{2}} | \uparrow_{\bf a}) & \propto & 
e^{-x^2}\{
1+e^{-2\alpha^2 (1-\eta)}\times
\nonumber \\
& &
\times [ a_x\cos (\sqrt{8\eta}\alpha x)+a_y\sin (\sqrt{8\eta}\alpha x) ]\}.
\nonumber \\
\end{eqnarray}
The above formula shows that the visibility of the interference fringes
is proportional to the factor $\exp[-2\alpha^2(1-\eta)]$. It is
exactly the same factor which appears in the expression for the Bell
combination given in Eq.~(\ref{Eq:Smaxapprox}). Thus, the better the
visibility of the interference fringes is, the stronger the violation of
Bell's inequalities takes place. If $\eta$ is interpreted
as the parameter describing interaction with a reservoir, it is
clearly seen that the violation of Bell's inequality and the 
interference visibility decay at the same rate.

When passing to the mesoscopic domain,
demonstration of quantum nonlocality in our scheme requires use of a
measuring apparatus which is capable of detecting interference between
the components of the superposition. However, this sensitivity
is important only in the homodyne measurements performed for
the phase $\theta=\pi/2$. The other half of homodyne measurements
can be in principle realized with a detector which does not have
microscopic sensitivity. This provides another example of
a situation described first by
Yurke and Stoler \cite{YurkStolPRL97}, who presented
a scheme for the violation of local realism employing a combination
of sensitive and insensitive detectors. Analogously to their proposal,
the discrimination between two amplitudes of coherent states is not
sensitive to the efficiency of the detector. The second type of
the measurement used by them is the determination of the
photon number parity, which requires single photon resolution.
In our case, the sensitive measurement has the form of homodyne
detection capable of resolving interference fringes in the 
superpositions $|\Psi_{\pm}\rangle
 = (|\alpha\rangle \pm |-\alpha\rangle) /
\sqrt{N_{\pm}}$. Let us note that the task of distinguishing
the states $|\Psi_{+}\rangle$ and $|\Psi_{-}\rangle$ in our
scheme could in principle also be performed by measuring the photon
number parity, as these two superpositions have non-zero occupation
probabilities only for even or odd Fock states, respectively.

\section{Experimental prospects}
\label{Sec:Experimental}

An important practical aspect of experimental schemes for testing
quantum nonlocality is their sensitivity to various imperfections
\cite{Eberhard}. The effect of non-ideal spin projection is relatively
straightforward to describe. If we assume that in a fraction of events
the measuring device returns a flipped value of the spin, then the 
ideal spin projections along a direction ${\bf a}$
are replaced by a generalized two-element
positive operator-valued measure:
\begin{eqnarray}
\hat{P}_\uparrow ({\bf a}) & = & 
\frac{1+\xi}{2} | \uparrow_{\bf a} \rangle \langle 
\uparrow_{\bf a} |
+
\frac{1-\xi}{2} | \downarrow_{\bf a} \rangle \langle
\downarrow_{\bf a} |
\nonumber \\
\hat{P}_\downarrow ({\bf a}) & = & 
\frac{1-\xi}{2} | \uparrow_{\bf a} \rangle \langle 
\uparrow_{\bf a} |
+
\frac{1+\xi}{2} | \downarrow_{\bf a} \rangle \langle
\downarrow_{\bf a} |
\end{eqnarray}
where $| \uparrow_{\bf a} \rangle$ and $| \downarrow_{\bf a} \rangle$ are
the eigenvectors of the operator ${\bf a} \cdot \hat{\mbox{\boldmath
$\sigma$}}$ and the parameter $\xi$, bounded between $0$ and $1$,
characterizes the efficiency of the spin measurement: for $\xi=1$ we
recover perfect spin projections, whereas $\xi = 0$ corresponds to the
completely noisy limit. A simple calculation shows that within such a
model of imperfect spin measurements each of the correlation functions
defined in Eq.~(\ref{Korel}) is multiplied by the parameter $\xi$.
Consequently, the complete Bell
combination $S$ becomes rescaled by the factor $\xi$ smaller
than one. Thus, in the limit $\sqrt{\eta}\alpha \gg 1$ we obtain:
\begin{equation}
S_{\text{max}} \approx 2 \xi \sqrt{1+
\left(\frac{2}{\pi}\exp[-2\alpha^2 (1-\eta)]\right)^2}.
\end{equation}

\begin{figure}
\includegraphics{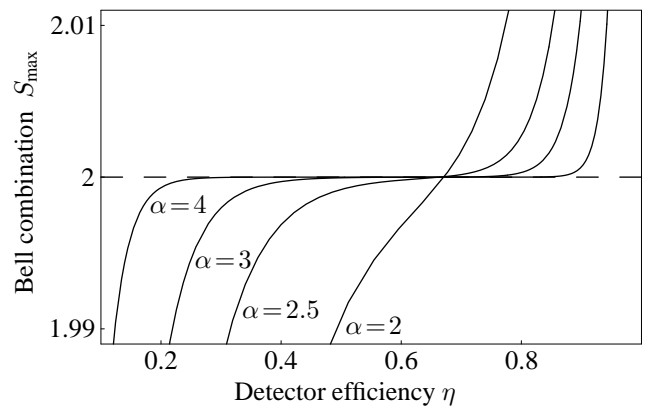}
\caption[]{\label{wykres3c} Maximum value of the Bell combination for
several values of $\alpha$ as a function of the detection efficiency
$\eta$, assuming perfect spin projections. For $\alpha = 2$ the minimum
efficiency necessary to demostrate nonlocality of the cat state is
slightly above 66\%.}
\end{figure}

For homodyne detection performed on the harmonic oscillator
subsystem, the role played by the efficiency parameter $\eta$ is more
involved. In Fig.~\ref{wykres3c} we plot the maximum value of the Bell
combination $S_{\text{max}}$ as a function of the homodyne detector
efficiency $\eta$ for several values of the coherent state amplitude
$\alpha$, under the assumption of perfect spin projections. According
to these numerical results, the lower bound for the homodyne detector
efficiency enabling the violation of Bell's inequality is about 66\%. 
We have found that this bound, clearly seen in the plot for $\alpha = 2$,
shifts to even slightly lower values with increasing $\alpha$. However,
for larger $\alpha$ the strict bound becomes rather meaningless, as in
its vicinity the violation of Bell's inequality becomes negligibly small.
Thus, its observation would require first huge sample of experimental data,
and secondly completely perfect spin measurements.

We shall close this section with a brief review of experimental prospects
for demonstrating the violation of Bell's inequalities using Schr\"{o}dinger
cat states. Among possible realizations, the physical systems which
would be most likely to achieve strict locality conditions include
a single radiation mode entangled with an atom \cite{CavityQED,Savage},
or with polarization states of a single photon \cite{Genovese}. The
radiation mode can either be confined in a high-$Q$ cavity or
travel in free space. In the latter case, the homodyne measurement is a
well-established technique used widely in quantum-optical experiments,
with the advantage of achieving very high detection efficiencies. In the
case of the measurement of quadratures for radiation fields in a cavity,
several schemes are available \cite{CavityHomodyne}, which remain however
considerably more complicated in practical realization.

\section{Conclusions}
\label{Sec:Conclusions}

In summary, we have shown how Schr\"{o}dinger cat states exhibiting
quantum entanglement can be applied to test the violation of Bell's
inequalities. The discussion was based on a concrete realization of the
scheme using the homodyne detection technique for a light mode. We have
analyzed the effects of losses and
imperfections, including the non-unit detection
efficiency, and pointed out a direct link between the visibility of
interference and the violation of Bell's inequalities. 
We have found that when passing to the macroscopic domain,
a substantial violation
of Bell's inequalities is possible in the limit
of perfect noise-free detection and absence of dissipation.
In a realistic case, losses destroy the nonlocal effects at the same
rate as they decrease the visibility of quantum interference between
the classically distinguishable wave packets in the superposition.

\section*{Acknowledgements}

We would like to acknowledge useful discussions with J. H. Eberly,
S. Wallentowitz, I. A. Walmsley, K. W\'{o}dkiewicz,
and M. \.{Z}ukowski. This research
was partially supported by ARO--administered MURI grant No.\
DAAG-19-99-1-0125, NSF grant PHY-9415583, and KBN grant 2~P03B~089~16.


\begin{thebibliography}{12}
\bibitem{Erwin}
E. Schr\"{o}dinger, 
Naturwissenschaften {\bf 23}, 807 (1935); {\bf 23}, 823 (1935); {\bf 23}, 
844 (1935).

\bibitem{Ion}
C. Monroe {\em et al.},
Science {\bf 272}, 1131 (1996).

\bibitem{CavityQED}
M. Brune {\em et al.}, 
Phys. Rev. Lett. {\bf 77}, 4887 (1996).

\bibitem{Koty}
B. Yurke and D. Stoler, Phys. Rev. Lett {\bf 57}, 13 (1986); S. Song,
C. M. Caves, and B. Yurke, Phys. Rev. A {\bf 41}, R5261 (1990);
V. Bu\v{z}ek, H. Moya-Cessa, P. L. Knight, and S. J. D. Phoenix, {\em
ibid.} {\bf 45}, 8190 (1992); I. A. Walmsley and M. G. Raymer, {\em ibid.}
{\bf 52}, 681 (1995); P. Tombesi and D. Vitali, Phys. Rev. Lett. {\bf 77},
411 (1996).

\bibitem{Savage}
C. M. Savage, S. L. Braunstein, and D. F. Walls, Opt. Lett. {\bf 15},
628 (1990)

\bibitem{Genovese}
M. Genovese and C. Novero, Phys. Rev. A {\bf 61}, 032102 (2000);
Phys. Lett. A {\bf 271}, 48 (2000)

\bibitem{YurkStolPRL97}
B. Yurke and D. Stoler, Phys. Rev. Lett. {\bf 79}, 4941 (1997).

\bibitem{HomodyneBell}
For use of homodyne detection in testing Bell's inequalities, see:
P. Grangier, M. J. Potasek, and B. Yurke, Phys. Rev. A {\bf 38},
R3132 (1998); S. M. Tan, D. F. Walls, and M. J. Collett, Phys. Rev.
Lett. {\bf 66}, 252 (1991); B. C. Sanders, Phys. Rev. A {\bf 45},
6811 (1992); {\bf 46}, 2966 (1992); A. Gilchrist, P. Deuar, and
M. D. Reid, Phys. Rev. Lett.  {\bf 80}, 3169 (1998); K. Banaszek and
K. W\'{o}dkiewicz, {\em ibid.} {\bf 82}, 2009 (1999); A. Kuzmich,
I. A. Walmsley, and L. Mandel, {\em ibid.} {\bf 85}, 1349 (2000).

\bibitem{WheelerZurek}
J. A. Wheeler and W. H. Zurek, {\em Quantum Theory and Measurement},
(Princeton University Press, Princeton, 1983).

\bibitem{WallMilbPRA85}
D. F. Walls and  G. J. Milburn, Phys. Rev. A {\bf 31}, 2403 (1985).

\bibitem{CatReview}
V. Bu\v{z}ek and P. L. Knight, in {\em Progress in Optics XXXIV}
ed. by E. Wolf (North-Holland, Amsterdam, 1995).

\bibitem{KW}
K. W\'{o}dkiewicz,
Opt. Comm. {\bf 179}, 215 (2000); New J. Phys. {\bf 2}, 21 (2000).

\bibitem{KiK}
S. L. Braunstein, Phys. Rev. A {\bf 42}, 474 (1990);
W. Vogel and J. Grabow, Phys. Rev. A {\bf 47}, 4227 (1993);
K. Banaszek and K. W\'{o}dkiewicz,
Phys. Rev. A {\bf 55}, 3117 (1997)

\bibitem{Bell}
J. F. Clauser, M. A. Horne, A. Shimony, and R. A Holt,
Phys. Rev. Lett. {\bf 23}, 880 (1969); J. S. Bell, in {\em Foundations
of Quantum Mechanics}, ed. by B. d'Espagnat (New York, Academic, 1971).

\bibitem{Eberhard}
 P. Eberhard, Phys. Rev. A {\bf 47}, R747 (1993)

\bibitem{CavityHomodyne}
H. M. Wiseman and G. J. Milburn, Phys. Rev. A {\bf 47}, 642 (1993);
M. Fortunato, P. Tombesi, and W. P. Schleich, Phys. Rev. A {\bf 59},
718 (1999).

\end{thebibliography}
\end{document}